\documentclass[%
 aps,pra,reprint,
 superscriptaddress, 
 showpacs,showkeys,
 longbibliography,noeprint,nopreprintnumbers,
 amsmath,amssymb,
 floatfix
]{revtex4-1}

\usepackage[pdftex]{graphicx}%  figure files
\usepackage[dvipsnames]{xcolor}

\usepackage{braket}
\usepackage{bm}% bold math
\usepackage{bbold} % black­board bold font
\usepackage{hyperref}% add hypertext capabilities

\newcommand{\hatA}{\ensuremath{\hat{A}}}
\newcommand{\hata}{\ensuremath{\hat{a}}}
\newcommand{\hatb}{\ensuremath{\hat{b}}}
\newcommand{\hatd}{\ensuremath{\hat{d}}}
\newcommand{\hatD}{\ensuremath{\hat{\mathcal{D}}}}
\newcommand{\hatP}{\ensuremath{\hat{\mathcal{P}}}}
\newcommand{\hatU}{\ensuremath{\hat{U}}}
\newcommand{\hatV}{\ensuremath{\hat{V}}}
\newcommand{\hamil}{\ensuremath{\hat{H}}}
\newcommand{\rhohat}{\ensuremath{\hat{\rho}}}
\newcommand{\identmat}{\ensuremath{\mathbb{1}}}
\newcommand{\nth}{\ensuremath{n^\mathrm{th}}}
\newcommand{\hatn}{\ensuremath{\rhohat}}

\begin{document}

\title{Measuring correlations of cold-atom systems using multiple quantum probes}

\author{Michael Streif}
 \email{michael.streif@physics.ox.ac.uk}
 \affiliation{Clarendon Laboratory, University of Oxford, Parks Road, Oxford OX1 3PU, United Kingdom}
 \affiliation{Physikalisches Institut, Albert-Ludwigs-Universit\"at Freiburg, Hermann-Herder-Stra{\ss}e 3, 79104 Freiburg, Germany}

\author{Andreas Buchleitner}
 \affiliation{Physikalisches Institut, Albert-Ludwigs-Universit\"at Freiburg, Hermann-Herder-Stra{\ss}e 3, 79104 Freiburg, Germany}

\author{Dieter Jaksch}
 \affiliation{Clarendon Laboratory, University of Oxford, Parks Road, Oxford OX1 3PU, United Kingdom}
\affiliation{Centre for Quantum Technologies, National University of Singapore, 3 Science Drive 2, 117543, Singapore}

\author{Jordi Mur-Petit}
 \email{jordi.murpetit@physics.ox.ac.uk}
 \affiliation{Clarendon Laboratory, University of Oxford, Parks Road, Oxford OX1 3PU, United Kingdom}

\date{10 October 2016}%\today}

\begin{abstract}
We present a non-destructive method to probe a complex quantum system using multiple impurity atoms as quantum probes.
Our protocol provides access to different equilibrium properties of the system by changing its coupling to the probes.
In particular, we show that measurements with two probes reveal the system's non-local two-point density correlations, for probe-system contact interactions.
We illustrate our findings with analytic and numerical calculations for the Bose-Hubbard model in the weakly and strongly-interacting regimes, under conditions relevant to ongoing experiments in cold atom systems.
\end{abstract}



\maketitle

\section{Introduction}

Different phases of matter are fundamentally associated with different correlations among their constituents. These correlations can be encoded in various observables. For example, the ground state of a one-dimensional, single-component Fermi gas has the same density profile as a one-dimensional system of strongly repulsive bosons (Tonks-Girardeau gas), while their momentum distributions are markedly different~\cite{Cazalilla2011}. This stems from the fact that the momentum distribution contains further information on the two-particle correlations, which also affect other observables such as the excitation spectrum and the structure factor of quantum systems~\cite{PinesNozieresVol1,Leggett2006book}.
While traditionally one could only access these properties via bulk measurements, e.g., neutron scattering off liquid helium, the advent of setups based on cold atoms in optical lattices has opened up new possibilities.
For example,
the measurement of local two-particle correlations in a one-dimensional gas of bosonic atoms for various interatomic (repulsive) interaction strengths was found~\cite{Kinoshita2005} to be in excellent agreement with theoretical calculations~\cite{Gangardt2003,Cazalilla2004,Kheruntsyan2005}.
Measurements of the momentum distribution~\cite{Paredes2004} and
non-local density-density correlation function~\cite{Folling2005}
of one-dimensional bosons in a periodic potential have also been performed, and they agree with theoretical findings~\cite{Paredes2004,Altman2004}.
More recently, $N_p$-point non-local correlation functions up to $N_p=10$ between two quasi-one-dimensional Bose gases, were measured by matter-wave interferometry~\cite{Langen2015}. These results 
underpin the necessity to account for conserved quantities in the description of the non-equilibrium evolution of 
quantum systems~\cite{Madronero2006,Rigol2008,Hiller2012,Birman2013,Freese2016}.

Common to all these experiments is that they use destructive measurements to study the quantum systems, most frequently the time-of-flight technique, where the trapping potential is switched off and the system allowed to expand before light absorption images are recorded.
Based on the development of new measurement and control methods, such as the quantum gas microscope (which enables access to quantum lattice systems with single-site resolution)~\cite{Bakr2009,Gemelke2009,Sherson2010,%
Ott2016,Kuhr2016}, an alternative approach is advancing which considers the use of other quantum objects, such as photons, single atoms or ions as non-destructive quantum probes of many-body quantum systems~\cite{Ponomarev2006,Mekhov2007,Sanders2010,Zipkes2010,
Schmid2010,Will2011,Hunn2012,Spethmann2012,Fukuhara2013,%
Mayer2014,Kozlowski2015}.

The idea of using single quantum probes---which often are equipped with the simplest possible internal quantum structure of a qubit---has been implemented to infer diverse properties of the host substrate, from Fr\"ohlich polarons, to work statistics and quantum phase transitions, to the Efimov effect and more \cite{Hohmann2015,Hohmann2016,Chin2010,%
Dorner2013,Mazzola2013,Batalhao2014,%
Gessner2014,Cosco2015,Johnson2016thermo,%
Elliott2016,Levinsen2015,Haikka2011}.
Yet it is clear that a single qubit probe in general cannot suffice to map out the host's characteristic properties exhaustively, since the probe-system coupling and the thus defined local density of states will generally limit the probe's diagnostic horizon to a finite subset of the system's Hilbert space.
It is therefore natural to seek a systematic generalization of the quantum probe approach to larger numbers of probes, such as to complement the finite diagnostic power of a single probe, e.g., by directly monitoring spatial correlations.

In the present contribution, we make a first step in this direction by considering two impurities embedded into a host bosonic gas~\cite{StreifMSc}.
Specifically, we show that the coherence of a two-probe density matrix enables us to access the two-point correlation function of a strongly-correlated quantum system in a non-destructive way.
We start in Sec.~\ref{sec:protocol} with a general presentation of our two-probe protocol. In Sec.~\ref{sec:numerics} we study a specific model of bosonic particles in a lattice, the Bose-Hubbard model (BHM), and show that our protocol enables us to determine the average system density as well as the two-point density-density correlation function, both in the superfluid and in the insulating phases of the BHM.
Finally, in Sec.~\ref{sec:con}, we conclude with a summary of our findings and an outlook.

%%%%%%%%%%%%%%%%%%%%%%%%%%%%%%%%%%%%%%%%%%%%%%%%%%%%%%%%%%%%%%%%%%%%
%%%%%%%%%%%%%%%%%%%%%%%%%%%%%%%%%%%%%%%%%%%%%%%%%%%%%%%%%%%%%%%%%%%%
%%%%%%%%%%%%%%%%%%%%%%%%%%%%%%%%%%%%%%%%%%%%%%%%%%%%%%%%%%%%%%%%%%%%

\section{Two-probe probing protocol}
\label{sec:protocol}

We consider a quantum system, $S$, coupled to two probes, which we label as $L$ (for left) and $R$ (right). 
The Hamiltonian of the composite system can be written as
\begin{align}
 \hamil_{\mathrm{tot}}
 =&
 \hamil_S \otimes\identmat_L \otimes\identmat_R
 +\identmat_S \otimes\hamil_L \otimes\identmat_R
 \nonumber \\
 &+\identmat_S \otimes\identmat_L \otimes\hamil_R
 +\hamil_\mathrm{int} \:,
 \label{eq:Htot}
\end{align}
where $\hamil_S$ is the Hamiltonian of the system and acts on the Hilbert space $\mathcal{H}_S$, $\hamil_{\alpha}$ ($\alpha=L,R$) is the Hamiltonian of the left (right) probe acting on its corresponding Hilbert space $\mathcal{H}_\alpha$, and $\hamil_\mathrm{int}$ is the interaction Hamiltonian between the system and the two impurities and therefore acts on $\mathcal{H}_\mathrm{tot} = \mathcal{H}_S \otimes\mathcal{H}_L \otimes\mathcal{H}_R$.

We model the probes as two-level systems (qubits), and couple them separately to the system, so that the interaction Hamiltonian reads
\begin{align}
 \hamil_\mathrm{int}
 &=
 \hatV_{SL} \otimes \bigg( g_{L0}\ket{0}_L\bra{0}_L
					+g_{L1}\ket{1}_L\bra{1}_L \bigg)
		  \otimes\identmat_R 
 \nonumber  \\
 &+
 \hatV_{SR} \otimes\identmat_L
 	\otimes\bigg( g_{R0}\ket{0}_R\bra{0}_R
 	+g_{R1}\ket{1}_R\bra{1}_R \bigg) .
 \label{eq:Hint}
\end{align}
Here, we have indicated the internal states of each probe qubit by $\ket{0}_\alpha,\ket{1}_\alpha$, respectively, and the parameters $g_{\alpha q}$ describe the interaction between the system and qubit $\alpha=L,R$ when in state $q=\ket{0},\ket{1}$.

Our probing protocol starts with the qubits uncoupled from the system, $g_{\alpha q}(t=0)=0$. The compound initial state reads $\rhohat_\mathrm{tot}(t=0)
=
\rhohat_S \otimes\ket{\Phi_{+}}\bra{\Phi_{+}}$, i.e., with the two qubits not entangled with the system, and prepared in the Bell state $\ket{\Phi_+} = \left( \ket{00}+\ket{11} \right)/\sqrt{2}$, with the usual notation $\ket{00} = \ket{0}_L\otimes\ket{0}_R$ and similarly for $\ket{11}$.
This entangled state can be prepared from both qubits initially in the ground state $\ket{0}$ and then subjected to a Hadamard gate acting on the left qubit followed by a controlled-\textsc{NOT} gate (with the left qubit as control and the right as target)~\cite{Nielsen1998}.

\begin{figure}[tb]
 \includegraphics[width=\columnwidth]{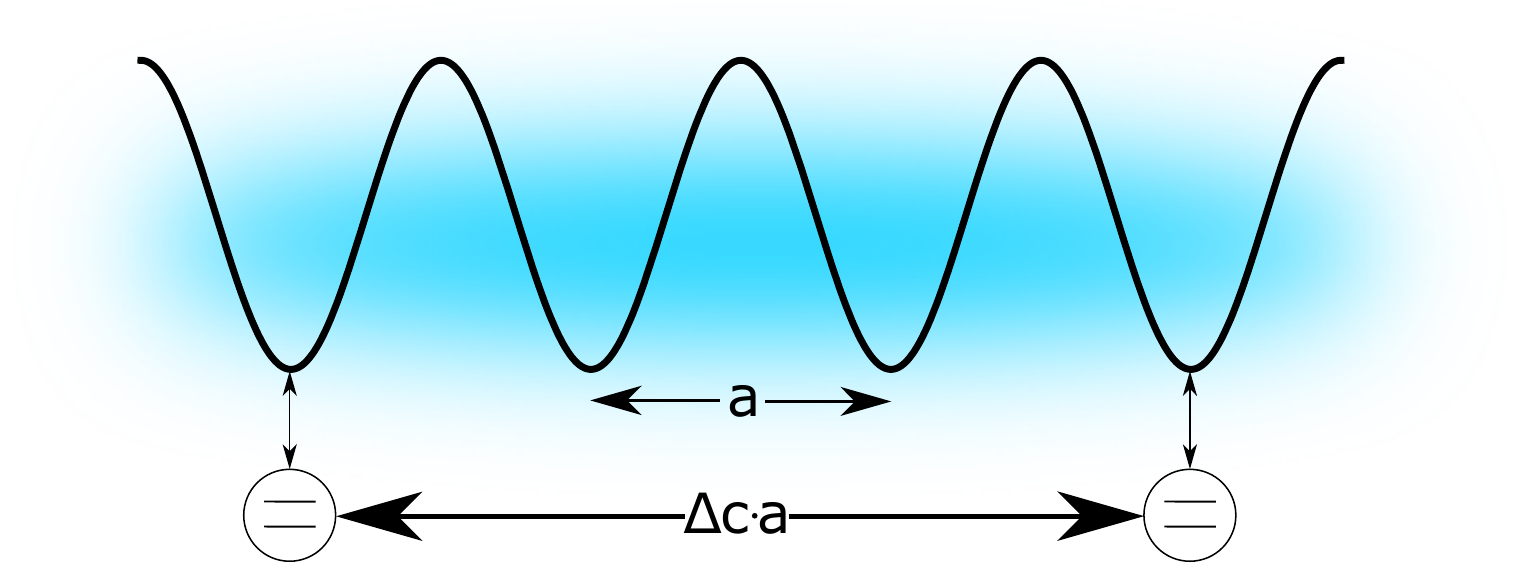}
 \caption{Schematic of a Bose gas (blue shading) in an optical lattice (black line). Two ancillary two-level quantum systems (circles) are coupled to the Bose gas at distinct sites on the lattice, separated by a distance $\Delta{c}$ in units of the lattice constant $a$.}
 \label{fig:scheme}
\end{figure}

At time $t=0$, a unitary non-equilibrium evolution is driven by changing the coupling of one of the internal states of the qubits with the system, e.g., by using a Feshbach resonance. For concreteness, we set $g_{L0}(t)=g_{R0}(t) \equiv g(t) = 1$ for $t>0$, while keeping
$g_{L1}(t)=g_{R1}(t)= 0$. The state of the composite system then evolves under the time evolution operator $\hatU(t)=\hat{\mathcal{T}} e^{-\frac{\mathrm{i}}{\hbar}\int_0^tdt' \hamil_\mathrm{tot}(t')}$, where $\hat{\mathcal{T}}$ is the time-ordering operator, so that after a time $t$ the composite system is in the state $\rhohat_\mathrm{tot}(t)=\hatU(t)\rhohat_\mathrm{tot}(0)\hatU^\dagger(t)$.
A trace over the system degrees of freedom yields the reduced density matrix operator of the two qubits, $\rhohat_\mathrm{Q}(t)=\mathrm{Tr}_S(\rhohat_{\mathrm{tot}}(t))$.
We focus our interest on the non-diagonal coherence element, whose time evolution can be expressed as $\braket{11|\rhohat_\mathrm{Q}(t)|00}=\frac{1}{4}\mathrm{e}^{-\frac{\mathrm{i}}{\hbar}t\Delta}\zeta(t)$.
Here, the exponential factor accounts for the free evolution in terms of the energy splitting between the internal states of the two probes, $\Delta=E_{\ket{11}}-E_{\ket{00}}$; without loss of generality, we set this energy difference to zero, i.e. $\Delta=0$.
The function $\zeta(t)$ characterizes the coherence element's time dependence due to the qubits' coupling to the system; we will refer to it as the coherence function.
Note that $\zeta(t)$ will generally depend on the distance between the probes, $\zeta(t)=\zeta(t; \Delta{c})$ (cf. Fig.~\ref{fig:scheme}), which we indicate explicitly where necessary.

The moments of the interaction Hamiltonian determine the derivatives of this coherence function~\citep{Elliott2016}. For example,
\begin{align}
 \frac{\mathrm{d}\zeta(t)}{dt}\bigg|_{t=0}
 &=\frac{\mathrm{i}}{\hbar} \braket{\hamil_{\mathrm{int}}} \, ,
 \label{eq:firstderiv}
 \\
 \frac{\mathrm{d^2}\zeta(t)}{\mathrm{dt^2}}\bigg|_{t=0}
 &= -\frac{1}{\hbar^2}\braket{\hamil_{\mathrm{int}}^2}\, ,
 \label{eq:secondderiv}
\end{align}
where the expectation values on the right hand sides are calculated with $\rhohat_\mathrm{tot}(t)$.
It follows that measurements of $\zeta(t)$ permit us to access several equilibrium expectation values of the system. 
These expectation values can be related to observables of interest by a suitable choice of the interaction between probes and system.
Below, we show that, in particular, for contact probe-system interactions, measurements of the coherence function provide a way to determine the density  [Eq.~\eqref{eq:first}] and the two-point density correlation function [Eq.~\eqref{eq:second}] of the host substrate.

Representing the internal state of each qubit as a spin operator, and using the Pauli spin matrices, $\sigma_i$ ($i=x,y,z$), the real and imaginary parts of $\zeta(t)$ can be written
\begin{align}
 \mathrm{Re}(\zeta(t))
 =&
 \frac{1}{2} \langle \hat{\sigma}_x \otimes\hat{\sigma}_x -\hat{\sigma}_y \otimes\hat{\sigma}_y \rangle_t
 \label{eq:Repart}\\
 \mathrm{Im}(\zeta(t))
 =&
 \frac{1}{2} \langle \hat{\sigma}_x \otimes\hat{\sigma}_y +\hat{\sigma}_y \otimes\hat{\sigma}_x \rangle_t,
\label{eq:Impart}
\end{align}
where the bracket $\langle \cdot \rangle_t$ represents a trace over $\rhohat_Q(t)$.
Thus, $\zeta(t)$ can be experimentally determined by measuring the two-qubit correlation functions which enter Eqs.~\eqref{eq:Repart} and \eqref{eq:Impart}.
Alternatively, one can express $\zeta(t)$ in the Bell basis as
$\mathrm{Re}(\zeta(t)) 
= 
2 \left( \rho_{Q,++}(t) - \rho_{Q,--}(t) \right)$,
$\mathrm{Im}(\zeta(t)) 
= 4\, \mathrm{Im} \left( \rho_{Q,+-}(t) \right)$, with
$\rho_{Q,++}(t) = \langle \Phi_+ | \rhohat_Q(t) | \Phi_+ \rangle$ and analogously for $\rho_{Q,--}$ and $\rho_{Q,+-}$, with $\ket{\Phi_-}=(\ket{00}-\ket{11})/\sqrt{2}$. It follows that $\zeta(t)$ can also be determined with Bell-state measurements.

%%%%%%%%%%%%%%%%%%%%%%%%%%%%%%%%%%%%%%%%%%%%%%%%%%%%%%%%%%%%%%%%%%%%
%%%%%%%%%%%%%%%%%%%%%%%%%%%%%%%%%%%%%%%%%%%%%%%%%%%%%%%%%%%%%%%%%%%%
%%%%%%%%%%%%%%%%%%%%%%%%%%%%%%%%%%%%%%%%%%%%%%%%%%%%%%%%%%%%%%%%%%%%

\section{Application to the Bose-Hubbard model}
\label{sec:numerics}

We now apply the protocol described in Sec.~\ref{sec:protocol} to the case of $N$ cold bosonic atoms loaded into the lowest energy band of an optical lattice with $M$ sites, described by the Bose-Hubbard Hamiltonian~\cite{Jaksch1998,Greiner2002}
\begin{align}
 \hamil_S
 &=
 -J\sum_{\braket{i,j}}\hata_i^\dagger\hata_j
 +\frac{U}{2} \sum_{i=1}^M\hata_i^\dagger\hata_i^\dagger\hata_i\hata_i
 +\mu \sum_{i=1}^M\hata_i^\dagger\hata_i
 \:.
 \label{eq:bosehubbard}
\end{align}
The operator $\hata_i^\dagger$ ($\hata_i$) creates (annihilates) a boson at a lattice site $i=1,\ldots,M$, the index $\braket{i,j}$ indicates summation over nearest neighbor pairs, and the parameters $U$, $J$, and $\mu$ are the on-site interaction energy, the hopping energy and the chemical potential, respectively.
We are interested in the translationally invariant system, i.e., in the limit $\{N\to\infty$, $M\to\infty\}$ with fixed average density $n=N/M$.

We now account for both probe impurities by a coupling mediated via a contact density-density interaction potential,
\begin{align}
 \hatV_{S\alpha}
 &=
 \int \mathrm{d}\bm{x} n_{\alpha}(\bm{x})
 	\hat{\Psi}^\dagger(\bm{x}) \hat{\Psi}(\bm{x}) \:,
 & \alpha=L,R
 \label{eq:coupling}
\end{align}
where $\hat{\Psi}(\bm{x})=\sum_j w_j(\bm{x}) \hat{a}_j$ is the bosonic field annihilation operator of the system, with 
$w_j(\bm{x})$ the lowest energy Wannier function at lattice site $j=1,\ldots,M$,
and $n_{\alpha}(\bm{x})$ the density of qubit $\alpha$ at position $\bm{x}$. Assuming that both impurities are strongly localized at distinct lattice sites ($j_L$ and $j_R$), we find that they interact with the Wannier function of that very site only. Thus, the interaction term can be written in terms of the boson number operators at these sites, $\hatV_{S\alpha}=\eta_{\alpha} \hata_{j_{\alpha}}^\dagger\hata_{j_{\alpha}}$, the parameter $\eta_{\alpha}=J\int\mathrm{d}\bm{x}|w_\alpha(\bm{x})|^2n_{\alpha}(\bm{x})$ being a measure of the interaction strength between the bosons and the qubit at site $j_\alpha$.
For simplicity, we assume that the local interaction strengths at both probe locations are identical, i.e., $\eta_L=\eta_R \equiv \eta$.

Substitution of Eq.~\eqref{eq:coupling} into Eq.~\eqref{eq:Hint}, together with Eq.~\eqref{eq:firstderiv}, yields the expectation value of the interaction's contribution to the total Hamiltonian which, due to the specific form of $V_{S\alpha}$, is equal to the bosonic density $\rhohat(j)=\hata^\dagger_j \hata_j$ at site $j$:
\begin{equation}
 2\overline{\rho}
 =
 \braket{\rhohat(j)+\rhohat(j+\Delta{c})}
 =
 \frac{\hbar}{\mathrm{i}  \eta}
 \frac{\mathrm{d}\zeta(t;\Delta{c})}{dt}\bigg|_{t=0},
 \label{eq:first}
\end{equation}
For the first equality, we used that, for translationally invariant systems, $\braket{\rhohat(j)}=\braket{\rhohat(j+\Delta{c})}\equiv\overline{\rho}$, with the integer $\Delta{c}=j_R-j_L$ the distance between the two qubits in units of the lattice constant $a$ (see Fig.~\ref{fig:scheme}).
(In an experiment, this can be accomplished by trapping the two qubits in a separate optical lattice formed by crossing two laser beams; the inter-qubit distance $\Delta{c}$ can then be precisely tuned by changing the angle between the propagation directions of the beams; see, e.g.,~\citep{Huckans2009}.)

Similarly, using Eq.~\eqref{eq:secondderiv}, we find the bosonic density-density correlation function
$\mathrm{Cor}(\Delta{c})=\braket{\rhohat(j)\rhohat(j+\Delta{c})}$ in terms of the qubits' coherence function:
\begin{equation}
 \braket{\left[ \rhohat(j)+\rhohat(j+\Delta{c}) \right]^2}
 = -\frac{\hbar^2}{\eta^2}
  \frac{\mathrm{d}^2\zeta(t;\Delta{c})}{dt^2}\bigg|_{t=0} \:.
 \label{eq:second}
\end{equation}
Again, given the system's translational 
invariance, $\braket{\rhohat(j)^2}=\braket{\rhohat(j+\Delta{c})^2}$, the last expression can be rewritten as
\begin{align}
 \mathrm{Cor}(\Delta{c})
 &=
 \frac{\hbar^2}{2\eta^2} \frac{\mathrm{d^2}}{dt^2}
	\bigg[\frac{1}{2}\zeta(t;\Delta c=0) - \zeta(t;\Delta c) \bigg]
 \bigg|_{t=0} .
 \label{eq:deccor}
\end{align}
This result implies that measurements of the qubits' coherence function $\zeta(t)$ provide access to the system's density-density correlation function. We remark that this result depends on the qubits-system coupling, Eq.~\eqref{eq:coupling}, but not on the specific form of the system Hamiltonian $\hamil_S$ beyond its translational invariance.
In the following sections, we assess the experimental feasibility of our protocol by simulating the outcome of the protocol in both the superfluid ($U/J\ll 1$) and the insulating ($U/J \gg 1$) phases of the one-dimensional Bose-Hubbard model, and comparing them with exact results for $\mathrm{Cor}(\Delta c)$ in both limits.

%%%%%%%%%%%%%%%%%%%%%%%%%%%%%%%%%%%%%%%%%%%%%%%%%%%%%%%%%%%%%%%%%%%%
%%%%%%%%%%%%%%%%%%%%%%%%%%%%%%%%%%%%%%%%%%%%%%%%%%%%%%%%%%%%%%%%%%%%
%%%%%%%%%%%%%%%%%%%%%%%%%%%%%%%%%%%%%%%%%%%%%%%%%%%%%%%%%%%%%%%%%%%%

\subsection{Weak interactions: Superfluid phase}

In the regime of weak interactions ($U/J\ll 1$), we can use Bogoliubov theory~\citep{PitaevskiiBook} to calculate both the coherence function $\zeta(t)$ and the density-density correlation function $\mathrm{Cor}(\Delta{c})$ analytically (see also~\cite{Mayer2014} and~\cite{Mayer2015}).
We start from the Bose-Hubbard Hamiltonian, Eq.~\eqref{eq:bosehubbard}, for a one-dimensional system of homogeneous density $n$.
We first transform the annihilation operators from the site basis, $\hata_i$, to the momentum basis, $\hatb_k=(M)^{-1/2}\sum_j \hata_j e^{\mathrm{i} k a j}$, and similarly for the creation operator $\hat{b}_k^\dagger$. A Bogoliubov transformation to quasiparticle operators, $\hatd_k=u_k\hatb_k+v_k\hatb_{-k}^\dagger$,
brings the system Hamiltonian into the diagonal form
$\hamil_S=\sum_k \hbar \omega_k \hatd_k^\dagger \hatd_k$,
with $\hatd_k$ ($\hatd_k^\dagger$) the annihilation (creation) operator of Bogoliubov quasiparticles of quasi-momentum $k$, and
$\omega_k=\sqrt{\epsilon_k(\epsilon_k+2U \overline{\rho})}$ the quasiparticle dispersion relation in terms of the single-particle energies $\epsilon_k=2J(1-\cos{(ka)})$, with $a$ the lattice constant and $ \overline{\rho}$ the bosonic density~\citep{LewensBook}.

With this transformation, we rewrite the density matrix of the lattice bosons by expressing the bosonic operators in terms of Bogoliubov quasiparticle operators
\begin{align}
 \rhohat_j
 &=\hata^\dagger_j\hata_j
 =
 \frac{1}{M}\sum_{k,k'} \hatb_k^\dagger \hatb_{k'}e^{\mathrm{i} kaj}e^{-\mathrm{i}k'aj}\nonumber
 \\
 &=
 \overline{\rho}
 + \frac{\sqrt{N_0}}{M}\sum_k\sqrt{\frac{\epsilon_k}{\omega_k}}
   \left( d_k^\dagger e^{\mathrm{i}kaj}+ d_k e^{-\mathrm{i}kaj} \right).
 \label{eq:rhoBogol}
\end{align}
In the second line, we have applied Bogoliubov's approximation, i.e., we assume that the occupation of $k \neq 0$ modes is small [$(N-N_0)/N \ll 1$], and neglect terms of quadratic (or higher) order in quasiparticle operators~\citep{PitaevskiiBook,LewensBook}.
By inserting Eq.~\eqref{eq:rhoBogol} into the definition of the two-point density correlation function, we reach the following analytic expression valid in the weakly-interacting limit
\begin{align}
\mathrm{Cor}(\Delta{c})
&= \overline{\rho}^2 +
   \frac{\overline{\rho}}{M}\sum_k\frac{\epsilon_k}{\omega_k}
	 (2\nth_k+1) \cos{(ka \Delta{c})} \:,
\label{eq:analyticalexp}
\end{align}
where we have evaluated the occupations of the Bogoliubov modes in a thermal state, $\braket{\hatd_k^\dagger\hatd_k}=\nth_k = 1/( e^{\beta \hbar \omega_k}-1 )$, with $\beta$ the inverse temperature, and we have dropped the anomalous averages $\braket{\hatd_k \hatd_{-k}}$, as they are negligible at the low temperatures where the Bogoliubov approximation applies~\cite{Griffin1996}.
At zero temperature ($\beta\to\infty$), Eq.~\eqref{eq:analyticalexp} satisfies the sum rule established in Ref.~\cite{Damski2015} for density-density correlations in the ground state, which 
re-expresses the sum rule relating the dynamic structure factor to the static structure factor, which in turn is sensitive to two-body interactions in bosonic lattice systems~\cite{Menotti2003}.
\begin{figure}
\includegraphics[width=1\columnwidth]{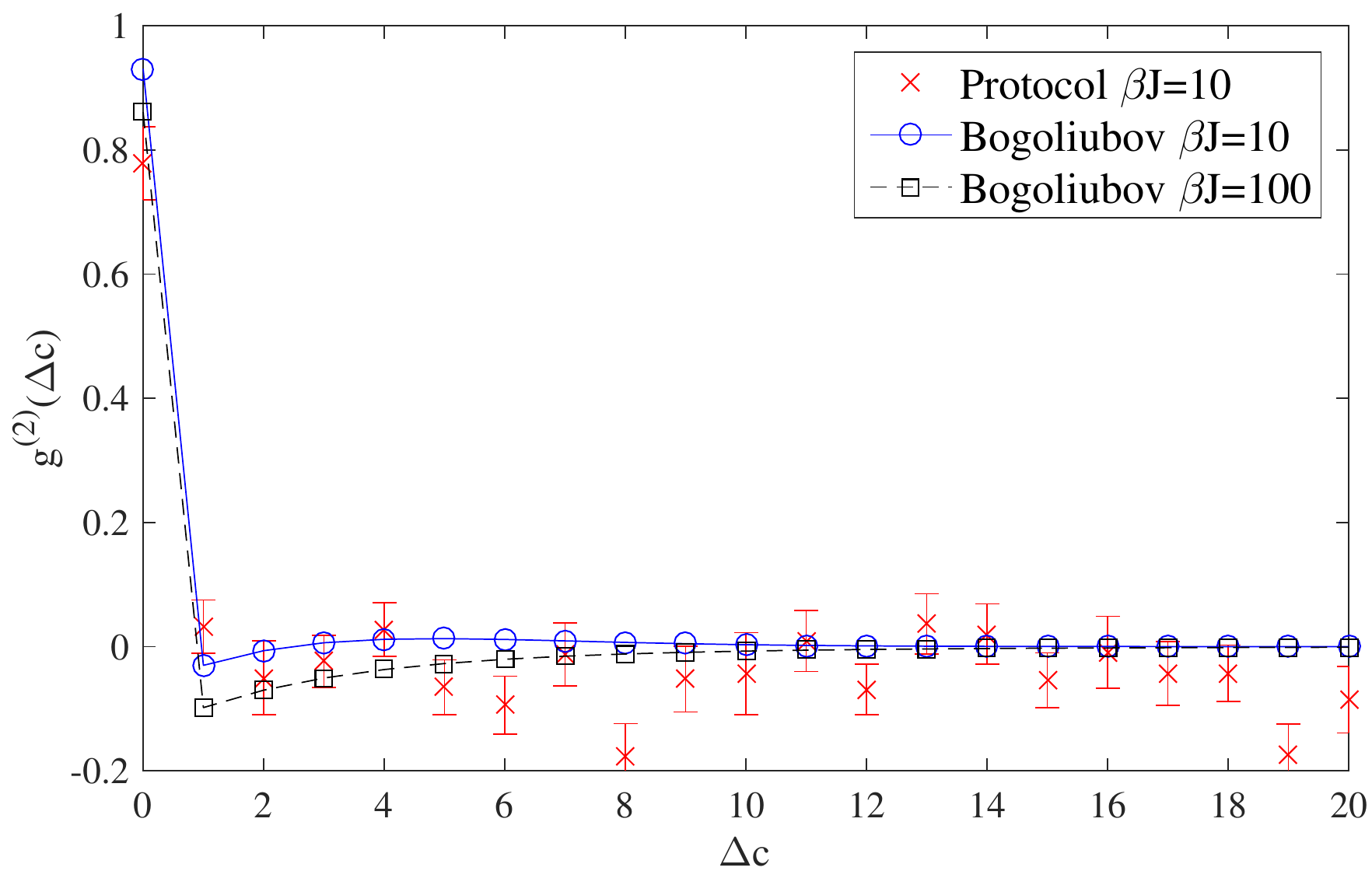}
 \caption{%
 \textit{Weak interactions.}
 Normalized correlation function obtained with Eq.~\eqref{eq:deccor} simulating $N_{\mathrm{exp}}=10^4$ experiments (crosses), compared with the analytic results Eq.~\eqref{eq:analyticalexp} (lines with symbols), for a system initially at equilibrium at inverse temperature $\beta J=10$ and $\beta J=100$. Other parameters used are $\eta=0.4J$ and $U/J=0.1$.}
 \label{fig:Bogosim}
\end{figure}

Based on Eq.~(\ref{eq:analyticalexp}), we plot in Fig.~\ref{fig:Bogosim} the normalized second-order correlation function,
\begin{align}
 g^{(2)}(\Delta{c})
 =\frac{ \braket{\rhohat(0)\rhohat(\Delta{c})}
		-\braket{\rhohat(0)}\braket{\rhohat(\Delta{c})} }
	   { \braket{\rhohat(0)}^2 },
 \label{eq:normalisedcorrelation}
\end{align} 
as a function of the inter-probe distance $\Delta{c}$ for different temperatures.
We see that, for all temperatures, the correlation vanishes for distances beyond a few lattice sites, which agrees with the picture that, in the non-interacting limit, the system is effectively described by a product of on-site coherent states so that
$\braket{\rhohat(0)\rhohat(\Delta{c})} = \braket{\rhohat(0)}\braket{\rhohat(\Delta{c})}$
\cite{Bloch2008rmp}. Weakly-interacting homogeneous one-dimensional Bose gases also converge to this limit fairly quickly~\cite{Deuar2009}.

We proceed now to compare these analytic calculations with the estimation by means of the coherence function $\zeta(t)$.
To evaluate the right-hand side of Eq.~\eqref{eq:deccor}, we rewrite the system-qubit interaction Hamiltonian in terms of Bogoliubov operators,
\begin{eqnarray}
 \hatV
 =
 \hatV_{SL} +\hatV_{SR}
 =
 2 \overline{\rho}\eta +\sum_k\left(\eta_k^\star
 \hatd_k^\dagger+\eta_k\hatd_k\right) ,
\end{eqnarray}
where
$\eta_k=\eta\sqrt{n\epsilon_k/M\omega_k}(e^{-\mathrm{i}kaj_L}+e^{-\mathrm{i}kaj_R} )$ is the coupling strength of the qubits with the Bogoliubov mode of quasi-momentum $k$.
Substituting these expressions into $\hamil_\mathrm{int}$ allows us to calculate analytically the time evolution of the composite system and, therefore, to determine the coherence function $\zeta(t)$.
Full details of the derivation are reported in Appendix~\ref{app:BogoApp} (see also~\cite{StreifMSc}); here we quote only the final result,
\begin{align}
 \zeta(t)
 &=
 e^{2\mathrm{i}\eta \overline{\rho} t}
 \exp \left[ -\mathrm{i} \sum_k \frac{|\eta_k|^2}{\omega_k^2}
 	[\omega_k t-\sin(\omega_k t) ] \right] 
 \nonumber \\
 &\times
  \exp{\left[ \sum_k \left( -2\frac{|\eta_k|^2}{\omega_k^2}
 	\sin^2 \frac{\omega_k t}{2}
 	\coth \frac{\beta\omega_k}{2} \right)  \right] }
 \:.
 \label{eq:decfuncbogo}
\end{align}

\begin{figure}[t]
 \includegraphics[width=\columnwidth]{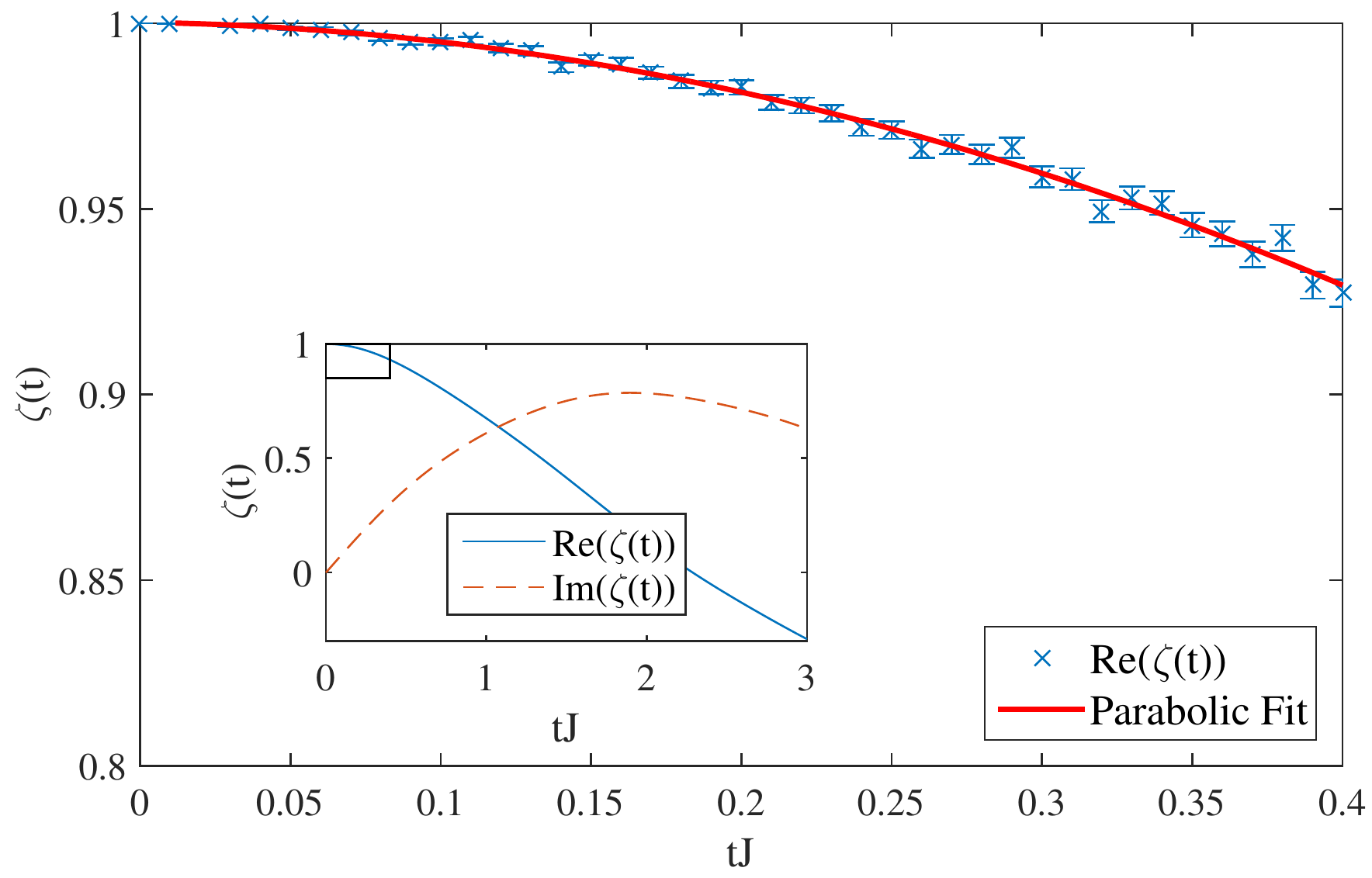}
 \caption{%
 Real part of the coherence function with added Gaussian noise (symbols) and a parabolic fit (solid line).
 Inset: Real (solid blue line) and imaginary (dashed red line) parts of $\zeta(t)$, Eq.~\eqref{eq:decfuncbogo}, for the case $\Delta{c}=5$. Here, we used a system with $M=1000$ lattice sites and $N=1000$ bosons; and, therefore, an average density $\overline{\rho}=1$. The system is initially in a thermal state with $\beta J=10$; other parameters as in Fig.~\ref{fig:Bogosim}.}
 \label{fig:decfunc}
\end{figure}
In an experiment, the coherence function $\zeta(t)$ can only be measured at discrete times, $t_r$.
In addition, for each time $t_r$, the expectation value defining $\zeta(t_r)$ is obtained upon accumulation of repeated measurements of the qubits' state, with individual measurement outcomes exhibiting quantum (shot) noise. To simulate this unavoidable spread of experimental measurement events, and to estimate how many measurements one would need for their statistical average to converge to the expectation value, we follow the scheme in Ref.~\cite{Johnson2016thermo} and add Gaussian noise to the calculated values of $\zeta(t_r)$; see Appendix~\ref{app:errors} for details on how to determine the corresponding variance.
As one would do in an experiment, to reduce the ensuing uncertainty in $\zeta(t_r)$, we repeat the simulated experiment a number $N_\mathrm{exp}$ of times and average over all outcomes, for each inter-probe distance $\Delta{c}$.
The values of $\zeta(t_r)$ estimated in this way are presented in Fig.~\ref{fig:decfunc} for a system with average density $\overline\rho=1$.
Here, one can note that the real part of $\zeta(t)$ has a parabolic dependence on time, while the imaginary part is linear around $t=0$.
It follows that the second derivative will be real, in accordance with our expectations for the density-density correlation function [cf.~Eq.~\eqref{eq:deccor}]. Thus, in practice it suffices to measure only the real part of $\zeta(t)$, Eq.~\eqref{eq:Repart}.

Given the smooth character of $\zeta(t_r)$, we fit a quadratic polynomial through these values, which enables us to calculate the right-hand side of Eq.~\eqref{eq:deccor} and determine the
two-point correlation function. 
We show the corresponding results for $\beta J=10$
in Fig.~\ref{fig:Bogosim}, which are in fair agreement with the analytic result~\eqref{eq:analyticalexp}. In particular, we see that the value of $g^{(2)}(0)$ derived from the protocol shows the characteristic enhancement of the superfluid phase. 
Reducing the statistical uncertainty of $g^{(2)}(\Delta{c})$ for $\Delta{c} \gg 1$ requires a relatively large number of measurements $N_\mathrm{exp}$, in line with previous experimental determinations of $g^{(2)}(\Delta{c})$ in cold atomic setups~\cite{Ottl2005,Schellekens2005}.
In the framework of the present two-probe protocol, these fluctuations, and correspondingly $N_\mathrm{exp}$, can be reduced by running in parallel an arrangement with $N_\mathrm{pairs}$ pairs of probes in a double-well superlattice~\citep{Anderlini2007,Folling2007,
Hofferberth2007,Hangleiter2015};
a setup with $N_\mathrm{pairs}=100$ probe pairs would reach the precision shown in Fig.~\ref{fig:Bogosim} with only 100 measurement runs.

%%%%%%%%%%%%%%%%%%%%%%%%%%%%%%%%%%%%%%%%%%%%%%%%%%%%%%%%%%%%%%%%%%%%
%%%%%%%%%%%%%%%%%%%%%%%%%%%%%%%%%%%%%%%%%%%%%%%%%%%%%%%%%%%%%%%%%%%%
%%%%%%%%%%%%%%%%%%%%%%%%%%%%%%%%%%%%%%%%%%%%%%%%%%%%%%%%%%%%%%%%%%%%

%
%
\begin{figure}[t]
 \includegraphics[width=\columnwidth]{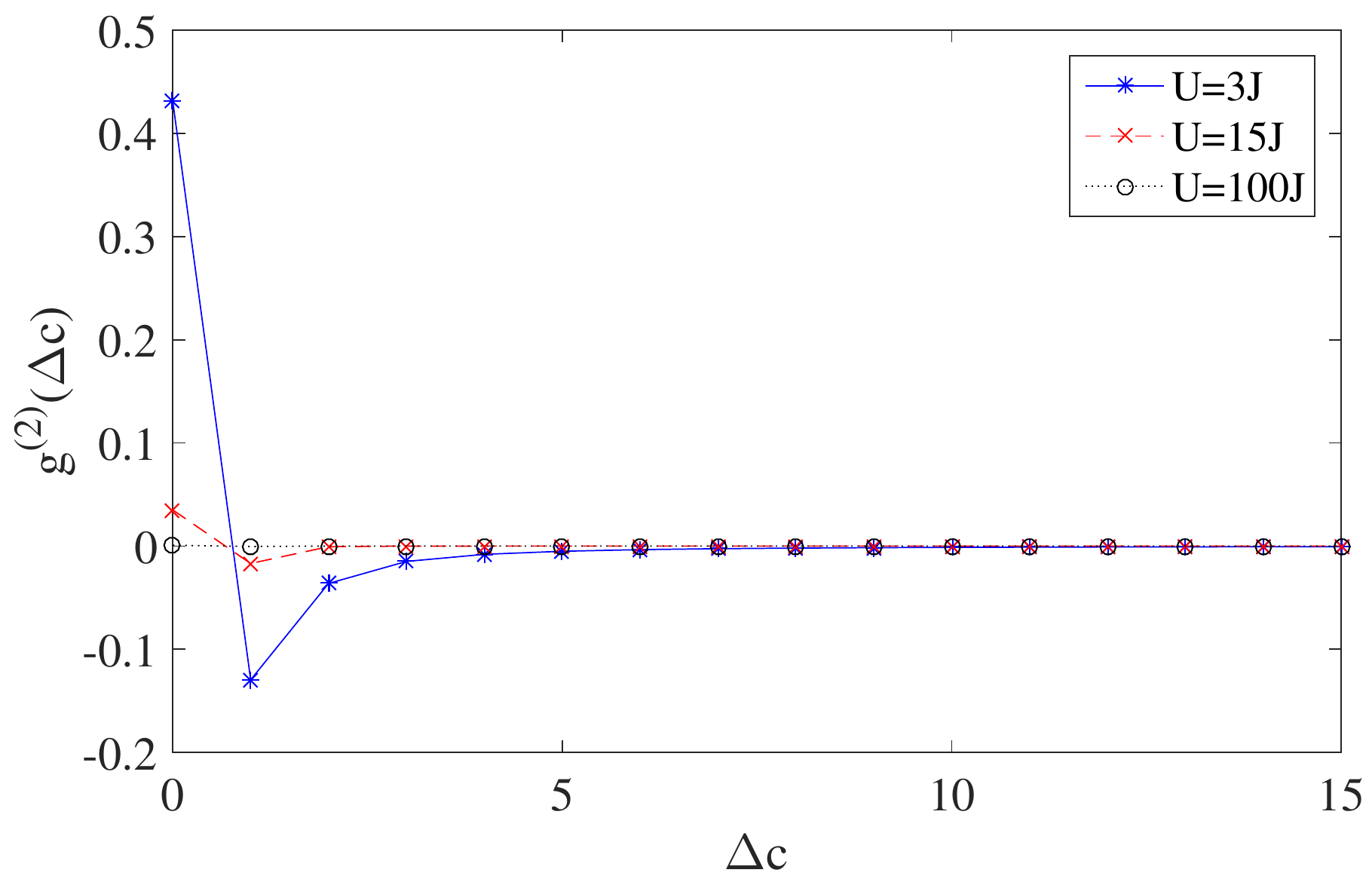}
 \caption{%
 Correlation function $g^{(2)}(\Delta{c})$ for the ground state in the strongly-interacting regime for different interactions strengths, $U/J=3$, $15$, and $100$, as indicated.
 These points represent the numerically exact expectation values of number operator pairs, $\left(\braket{\hatn_i \hatn_j}-\braket{\hatn_i}\braket{\hatn_j}\right)/\braket{\hatn_i}^2$, from the TNT calculation.
 }
 \label{fig:TNTcor}
\end{figure}

\subsection{Strong interactions: Insulating phase}

For stronger interactions $U/J \gtrsim 1$, the correlations between the bosons in the lattice invalidate an approach based on the Bogoliubov treatment. 
An efficient method to deal with this situation is Tensor Network Theory (TNT), which provides numerically exact ground state properties of strongly-correlated systems, in particular, of the one-dimensional BHM~\cite{Orus2014,TNTbook}. Here, we apply this method to calculate $g^{(2)}(\Delta{c})$ in the ground state of this model using the implementation Oxford TNT library~\cite{TNTweb}.
As we are interested in investigating non-local correlation functions, we choose a large system with $M=101$ lattice sites, and $\overline{\rho}=1$ as before, and calculate $g^{(2)}(\Delta{c})$ around the central lattice site so that boundary effects are negligible and the system can still be considered (approximately) translationally invariant.
For the calculations presented below, we have checked that sufficient accuracy is reached bounding the site occupation to a maximum of four bosons per site and fixing a truncation parameter (maximum number of Schmidt coefficients) of $\chi=100$.

The TNT method allows us to calculate directly the expectation values of the number operator at each lattice site, $\braket{\hatn_i}$, and all pairs of number operators, $\braket{\hatn_i \hatn_j}$. From these, we obtain directly the normalized two-point correlation function
$g^{(2)}(\Delta{c})
=
\left(
 \braket{\hatn_i \hatn_{i+\Delta{c}}}
-\braket{\hatn_i}\braket{\hatn_{i+\Delta{c}}}
\right)/\braket{\hatn_i}^2$; the results for increasing values of $U/J$ are shown in Fig.~\ref{fig:TNTcor}.
As expected, in the limit $U/J \rightarrow \infty$ we recover that $g^{(2)}(\Delta{c})=0$ $\forall\Delta{c}$ as the ground state is a product of on-site Fock states with no density fluctuations~\cite{Bloch2008rmp,Endres2013}.
These results constitute the test-bed corresponding to the left-hand site of Eq.~\eqref{eq:deccor}, which we will compare to the outcome of the protocol to obtain $\zeta(t)$ and its derivatives.

\begin{figure}[t]
 (a)~\raisebox{\dimexpr-\height+\baselineskip}{%
  \includegraphics[width=0.95\columnwidth]{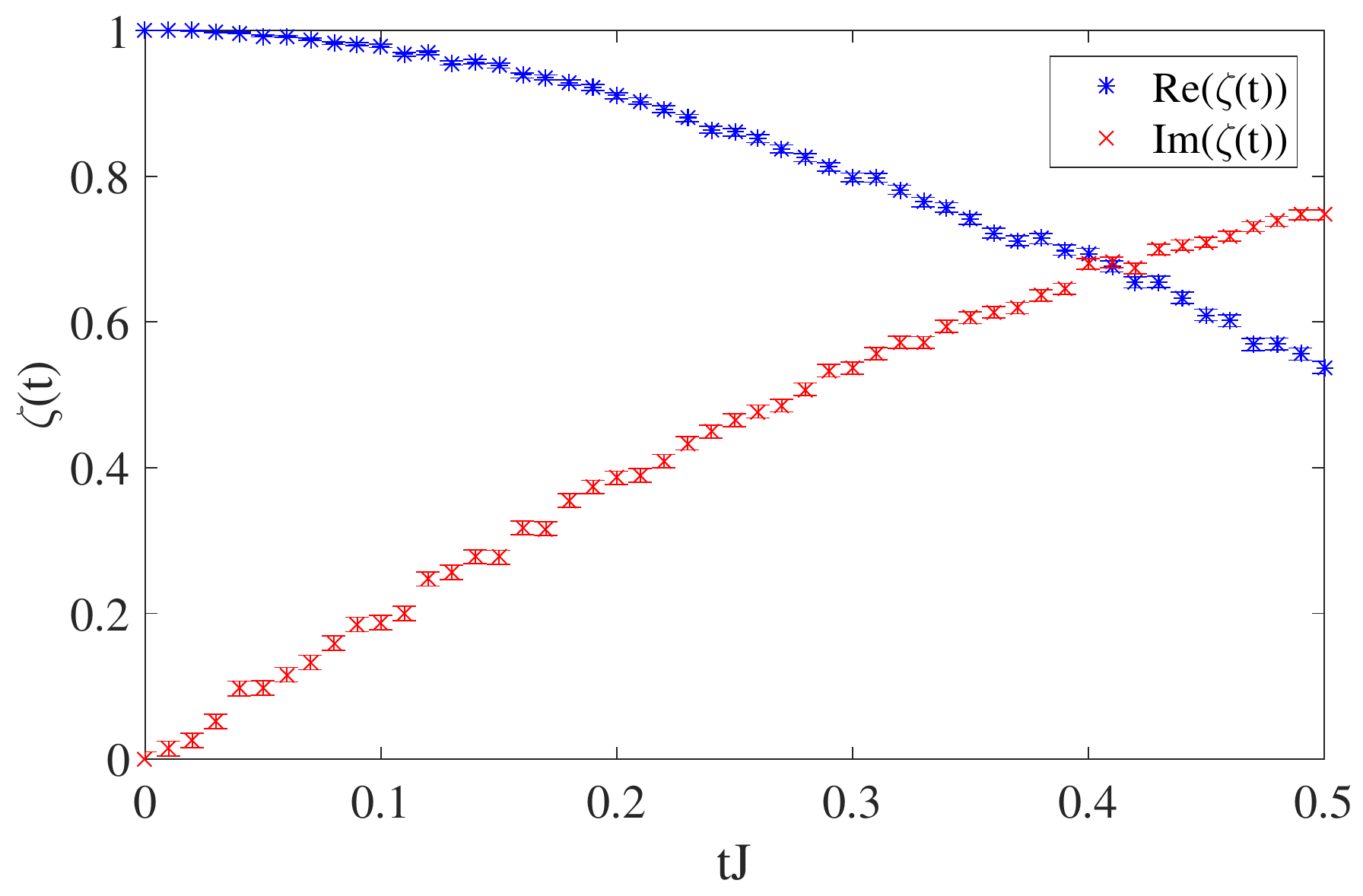}
  }
 (b)~\raisebox{\dimexpr-\height+\baselineskip}{%
  \includegraphics[width=0.95\columnwidth]{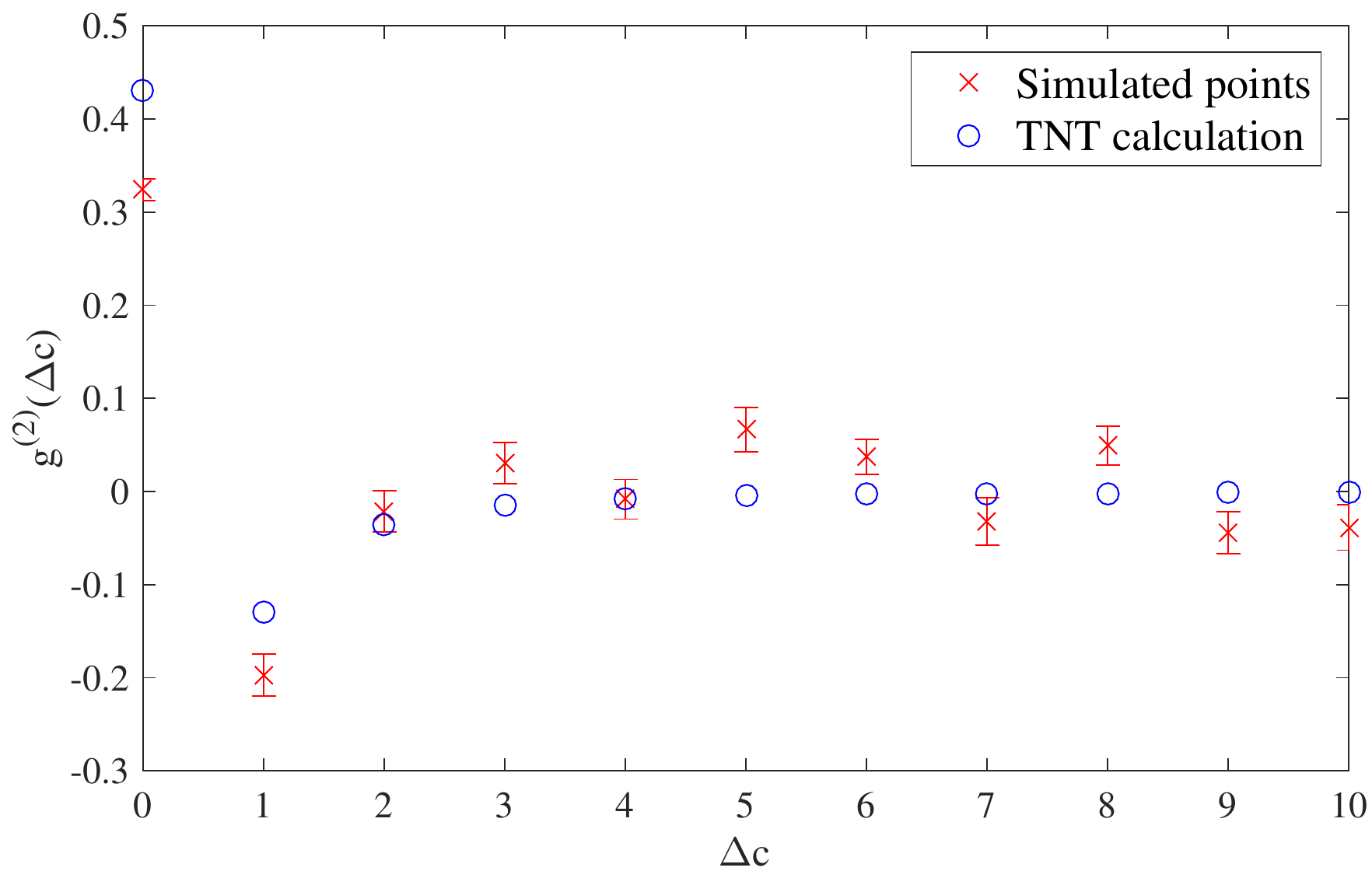}
  }
 \caption{%
 \textit{Strong interactions.}
 (a) Real (asterisks) and imaginary (crosses) parts of the coherence function $\zeta(t)$ for $U/J=3$. Other parameters are $\Delta{c}=5$, $M=101$, $\overline{\rho}=1$, and $\eta=J$.
 (b) Normalized correlation function $g^{(2)}(\Delta{c})$ for the same parameters. The results of the measurement protocol with $N_{\mathrm{exp}}=10^4$ (crosses) agree with the numerically exact values calculated with the TNT method (circles).
  }
  \label{fig:TNTsim}
\end{figure}
We calculate $\zeta(t)$ in the strongly-interacting regime in the following way:
The coherence function can be written as a trace over system operators only, $\zeta(t) = \mathrm{Tr}_S(\hatU_1(t) \hatU_0(t) \rhohat_\beta)$ [cf. Eq.~\eqref{eq:alternativezeta}].
Here, $\hatU_0(t)$ is the evolution operator over a time $t$ with the initial system Hamiltonian, while $\hatU_1(t)$ is the evolution operator including the coupling to the qubits.
For probe qubits localized at lattice sites and coupled to the bosons by contact interactions of strength $\eta$, the effect of the probe-boson coupling amounts to a local shift of the bosons' chemical potential, $\mu \to \mu-\eta$, at the sites where the probes are located.
Thus, we can obtain $\zeta(t_r)$ at different time steps $t_r$
by calculating the expectation value 
$\mathrm{Tr}_S(\hatU_1(t) \hatU_0(t) \rhohat_\beta)$ with $\rhohat_\beta$ the ground state of the bosonic system in a lattice with modified local potential at the probe sites.

\begin{figure}
  \includegraphics[width=1\columnwidth]{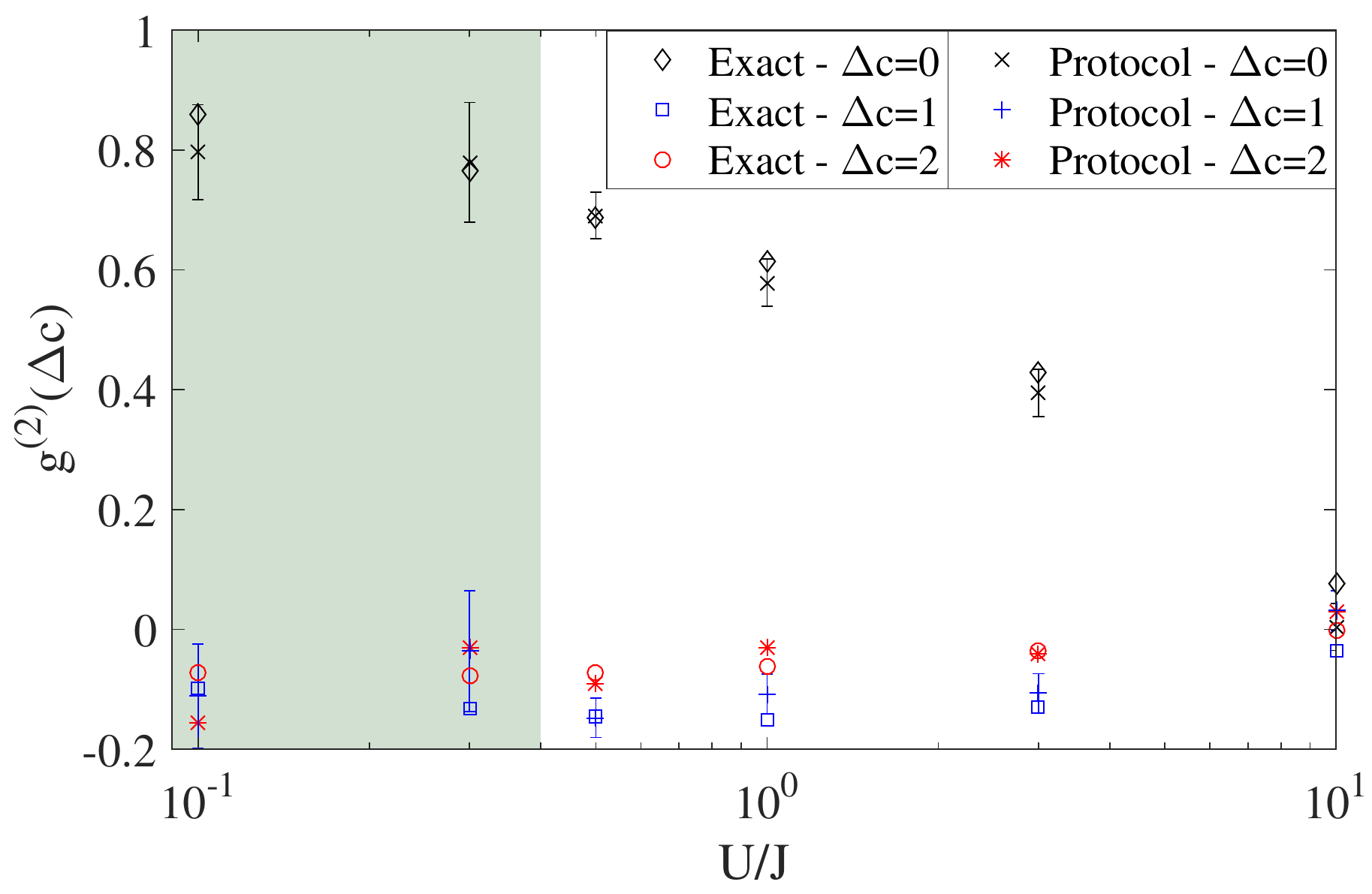}
  \caption{Correlation function $g^{(2)}(\Delta{c})$ for $\Delta{c}\in \{0,1,2\}$ and different values of $U/J$. Bogoliubov theory was used for $U/J\le 0.4$ with a $\beta J=1000$ thermal state (shaded region), and TNT for larger values of $U/J$. For clarity, we do not include error bars for the $\Delta{c}=2$ calculation; they are similar to those for $\Delta{c}=1$.}
  \label{fig:UJplot}
\end{figure}

In our numerical calculations, we take $t_r=r\Delta{t}$ with $\Delta{t}=\hbar\times 0.01/J$ and $r=0,\ldots,20$.
As for the weakly-interacting regime, we simulate the uncertainty in an experiment by adding noise to each simulated data point, $\zeta(t_r)$, and calculate the numerical second derivative at $t=0$. We repeat this procedure for all integer distances between the two qubits $0 \leq \Delta{c} \leq 15$ ($\ll M$ to avoid boundary effects).
The coherence function, $\zeta(t)$ obtained in this way is shown in Fig.~\ref{fig:TNTsim}(a).
We observe that both real and imaginary parts exhibit broadly a behavior similar to that of the weakly interacting system.
However, the correlation function that one obtains from this according to Eq.~(\ref{eq:deccor}) is notably different, as shown in Fig.~\ref{fig:TNTsim}(b), where we compare the value of $g^{(2)}(\Delta{c})$ obtained from the coherence function by using Eq.~\eqref{eq:deccor} with the numerically exact values derived from the TNT ground state (the latter values are the same as those in Fig.~\ref{fig:TNTcor} for $U/J=3$). We see that there is a good agreement between the two calculations, as happened in the weakly interacting regime. In particular, the estimation of the correlation function using our protocol is able to detect the reduction in $g^{(2)}(0)$ as the system gets deeper into the Mott insulating phase, $U/J \gg 1$.
To illustrate this point, we show in Fig.~\ref{fig:UJplot} the normalized correlation function $g^{(2)}(\Delta{c})$ at selected distances $\Delta{c}$ for different values of $U/J$ across the Mott insulator--to--superfluid transition.
First, we observe that the outcome of our protocol in each case is very close to the exact result (calculated with Bogoliubov theory for weak interactions and with TNT for stronger interactions).
Physically, the local correlation, $g^{(2)}(0)$, decreases steadily as the repulsion between bosons increases, and it vanishes in the limit $U/J \gg 1$.
Correlations at larger distances are negative (meaning, it is \textit{less} probable to find a particle at distance $\Delta{c}$ in the actual ground state than what one would predict by relying only on the average density) and generally of smaller magnitude than the local correlation; they also vanish in the strongly repulsive limit, as expected for a Mott insulator.

%%%%%%%%%%%%%%%%%%%%%%%%%%%%%%%%%%%%%%%%%%%%%%%%%%%%%%%%%%%%%%%%%%%%
%%%%%%%%%%%%%%%%%%%%%%%%%%%%%%%%%%%%%%%%%%%%%%%%%%%%%%%%%%%%%%%%%%%%
%%%%%%%%%%%%%%%%%%%%%%%%%%%%%%%%%%%%%%%%%%%%%%%%%%%%%%%%%%%%%%%%%%%%

\section{Discussion and Outlook}
\label{sec:con}

In this paper, we have developed a framework to study correlation functions in cold atom systems by using multiple atomic impurities as quantum probes, a setup realized in recent experiments where potassium~\cite{Ospelkaus2006a,Will2011,Catani2012} or cesium~\cite{Spethmann2012,Hohmann2016} atomic impurities
were immersed in larger rubidium Bose gases.

We have presented a protocol which is able to measure the density-density correlations of the system relying on measuring the internal states of two probes and studying an off-diagonal element of their reduced density matrix. We have shown that the results of this protocol agree with those of analytic and numerically exact calculations for a one-dimensional Bose-Hubbard model in both the weakly and the strongly interacting regimes.
In particular, we have shown that the protocol is able to witness the change in correlations across the superfluid--to--Mott insulator transition.

Non-local density correlations in quantum gases have previously been measured by various methods, including noise interferometry, Bragg spectroscopy, and matter-wave interferometry. Let us briefly contrast our proposal with these techniques.
In Bragg spectroscopy, some of the atoms in the system are excited by two-photon Bragg scattering into a state of given momentum and energy. This provides access to the dynamic structure factor of the gas, which is the Fourier transform of the density correlation function~\cite{Kozuma1999,Stenger1999,Menotti2003,Rey2005}.
This method inevitably destroys the initial quantum state of the system, in contrast to our proposal, which is inherently non-destructive and, thus, could permit a time-dependent monitoring of the evolution of correlations. In addition, our protocol can be extended to using $N>2$ quantum probes to determine $N$-point correlation functions.

Matter-wave interferometry~\cite{Cronin2009}
is a destructive measurement method especially suited to probing the phase structure of bosonic quantum gases.
As mentioned earlier, it has been used recently to measure density correlation functions up to $10^\mathrm{th}$ order between two quasi-one-dimensional bosonic gases~\cite{Langen2015}.
However, the application of this method to higher-dimensional systems would require a rather involved analysis of the corresponding multi-dimensional phase interference pattern.
In contrast, it is straightforward to see that our protocol applies to systems of any dimensionality.

Noise interferometry retrieves information on particle correlations in atomic gases by analyzing the shot-to-shot fluctuations in absorption images of the system after time-of-flight evolution~\cite{Altman2004,Greiner2005,Folling2005}.
In strongly correlated phases, where the time-of-flight technique is not suitable, one could implement noise interferometry by imaging the atoms with a quantum gas microscope~\cite{Bakr2009,Gemelke2009,Sherson2010,Ott2016,Kuhr2016,%
Cheuk2015,Haller2015,Miranda2015,Edge2015,%
Cocchi2016,Boll2016,Parsons2016,Drewes2016} to analyze correlations in optical lattice setups.
Our proposal constitutes a complementary approach of similar experimental complexity, particularly suited to multicomponent setups with impurities~\cite{Ospelkaus2006a,Will2011,%
Catani2012,Spethmann2012,Hohmann2016}, with the distinctive feature of allowing non-destructive measurements.

The main challenge of our proposal may lie in the dynamical control of the probe-system coupling. Manipulation via Feshbach resonances is an option if these are available between the atomic species involved.
More generally, one could envisage probing a one- or two-dimensional gas by allowing the impurities to ``fall through'' it, pulled by gravity or driven by an external field. This would turn on and off the interactions without changing the state of the system appreciably (given that there are many more atoms in the background than impurities). This approach could be implemented exploiting existing experimental schemes in which the impurities are trapped near but outside the system and then driven into it for a fixed amount of time~\cite{Spethmann2012,Hohmann2016}, or made to penetrate it periodically~\cite{Eto2016}.

In summary, the framework presented here opens up new possibilities for the experimental investigation of quantum many-body systems and, especially, systems of cold atoms in optical lattices.
The protocol can be extended in various ways, e.g., to estimate $N$-point correlation functions.
Another possibility stems from the freedom of choosing the kind of interaction Hamiltonian between qubits and system, different choices allowing one to gain access to different observables. For example, by using Raman transitions~\citep{Recati2005}, the evolution of the probes becomes sensitive to the phase of the matter wave and one could measure cross-correlation functions~\citep{Kaminishi2015}.

\begin{acknowledgments}
The authors would like to thank J. J. Mendoza-Arenas, T. H. Johnson, and M. Mitchison for useful discussions.
This work was supported by the
EU H2020 FET Collaborative project QuProCS (Grant Agreement No.\ 641277),
EU Seventh Framework Programme (FP7/2007-2013) Grant Agreement No.\ 319286 Q-MAC, 
and Erasmus Placements (M.S.).
D. J. thanks the Graduate School of Excellence Material Science in Mainz for hospitality during part of this work.
\end{acknowledgments}

\appendix

%%%%%%%%%%%%%%%%%%%%%%%%%%%%%%%%%%%%%%%%%%%%%%%%%%%%%%%%%%%%%%%%%%%%
%%%%%%%%%%%%%%%%%%%%%%%%%%%%%%%%%%%%%%%%%%%%%%%%%%%%%%%%%%%%%%%%%%%%

\section{Bogoliubov treatment of the weakly interacting system}
\label{app:BogoApp}
We briefly expand on the explicit calculation of the coherence function $\zeta(t)$ for weak interactions, with the help of Bogoliubov theory, and following the procedure outlined in \cite{Johnson2016thermo}. The first step is to introduce a set of projection operators on the Hilbert space of the two qubits,
\begin{align*}
 \begin{array}{ll}
  \hatP_{11}=\ket{11}\bra{11} \,,
  & ~~~
  \hatP_{10}=\ket{10}\bra{10} \,, \\
  \hatP_{01}=\ket{01}\bra{01} \,,
  & ~~~
  \hatP_{00}=\ket{00}\bra{00} \,.
 \end{array}
\end{align*}
This enables us to rewrite the full Hamiltonian in a more convenient form.
\begin{align*}
 \hamil_\mathrm{tot}
 &=
 \hatP_{11} \otimes\left(E_1 +\hamil_S+g_{L1}\hatV_{SL}
 	+g_{R1}\hatV_{SR} \right)
 + \\
 &+
 \hatP_{10} \otimes\left(E_2 +\hamil_S+g_{L1}\hatV_{SL}
    +g_{R0}\hatV_{SR}\right)
 + \\
 &+
 \hatP_{01} \otimes\left(E_3 +\hamil_S+g_{L0}\hatV_{SL}
    +g_{R1}\hatV_{SR}\right)
 + \\
 &+
 \hatP_{00} \otimes\left(E_4 +\hamil_S+g_{L0}\hatV_{SL}
    +g_{R0}\hatV_{SR}\right) \,
\end{align*}
As stated in the text, we are interested in the time evolution of the qubits only. Therefore, after calculating the time evolution of the composite system, we trace out the degrees of freedom of the bosons. After that, we concentrate on the coherence element of the two-qubit density matrix, $\braket{11|\rhohat_\mathrm{Q}|00}$. We find that the coherence function can be determined by calculating the expectation value 
\begin{align}
 \zeta(t)=\mathrm{Tr}_S(\hatU_1(t) \hatU_0(t) \rhohat_S)
 \label{eq:alternativezeta}
\end{align}
with the initial state of the system $\rhohat_S$.
In this expression
\begin{align*}
 \hatU_0(t) = \hat{\mathcal{T}}
 \exp{\left(-\frac{\mathrm{i}}{\hbar}
    \int_0^t \mathrm{dt'}\hamil_S\right)}
\end{align*}
is the time evolution operator with the unperturbed system Hamiltonian, and 
\begin{align*}
 \hatU_1(t) = \hat{\mathcal{T}}
 \exp{\left(-\frac{\mathrm{i}}{\hbar}
    \int_0^t \mathrm{dt'}(\hamil_S +g(\hatV_{SL}
 	+\hatV_{SR} ))\right)}
\end{align*}
is the time evolution operator with the Hamiltonian including the coupling to the probes, where we have used that $g_{L0}=g_{R0}=1$ and $g_{L1}=g_{R1}=0$.
It is worth noting the similarity of $\zeta(t)$ to the Loschmidt echo~\citep{Peres1984,Jalabert2001}, which is a function that enables us to characterize memory effects in the dynamics of quantum systems (see, e.g.,~\cite{Haikka2012}).

For simplicity, we change into the interaction picture, where $\hatU_0=\identmat$. The remaining time evolution operator simplifies to a more convenient expression:
\begin{align*}
 \hatU_1(t)
 =
 \hat{\mathcal{T}} \exp{ \left(-\frac{\mathrm{i}}{\hbar}
 	\int_0^t dt' \hatV_\mathrm{int}(t') \right)} \:.
\end{align*}
Here, $\hatV_\mathrm{int}(t)$ is the interaction part of the Hamiltonian in the interaction picture,
\begin{align*}
 \hatV_\mathrm{int}(t)
 =
 2 \overline{\rho}\eta +\sum_k\left( \eta_k^\star e^{\mathrm{i}\omega_kt}
  \hatb_k^\dagger +\eta_k\hatb_k e^{-\mathrm{i}\omega_kt} \right) \:.
\end{align*}
We can simplify the expression for $\hatU_1$ by applying the Magnus expansion~\cite{dattoli1986time}.
To this end, we introduce an operator $\hatA$ by
\begin{align*}
 \hat{\mathcal{T}} \exp{\left(-\frac{\mathrm{i}}{\hbar}
   \int_0^t dt' \hatV_\mathrm{int}(t') \right)}
   =\mathrm{e}^{\hatA}
 \:.
\end{align*}
This operator can be expressed as a sum of operators $\hatA=\sum_i \hatA_i$ which are related to commutators of the interaction Hamiltonian:
\begin{align*}
 \hatA_1 &= -\mathrm{i}\int_0^t dt' \hatV_\mathrm{int}(t')
 \\
 \hatA_2 &= \frac{1}{2}\int_0^t\mathrm{dt'}\int_{0}^{t'}\mathrm{dt''}
 	[\hatV_\mathrm{int}(t'),\hatV_\mathrm{int}(t'')]
 \\
 & \vdots .
\end{align*}
Given the form of $\hatV_\mathrm{int}$ above, the commutators at different times are c-numbers, $[\hatV_\mathrm{int}(t'),\hatV_\mathrm{int}(t'')]=-2i\sum_k|\eta_k|^2\sin{(\omega_k(t'-t''))}$. Therefore, all terms of the expansion beyond the second term vanish. Thus, we can write the coherence function as
\begin{align}
\zeta(t)
 &= e^{2\mathrm{i} \overline{\rho}\eta t}
 	\exp{\left[ -\mathrm{i}\sum_k\frac{|\eta_k|^2}{\omega_k^2}
 		[\omega_k t-\sin{(\omega_k t)}] \right] }
 \nonumber
 \\
 &\times
 \mathrm{Tr}
 \left[\exp\left\{ -\mathrm{i}\sum_k
	 	\left(\gamma_k\hatd_k^\dagger +\gamma_k^\star \hatd_k\right)
 	\right\} \rhohat_S \right]
 \label{eq:zetamagnus}
\end{align}
where we have defined
$\gamma_k = \frac{\eta_k^\star}{\omega_k} \left(\frac{e^{i\omega_kt}-1}{\mathrm{i}}\right)$.
We are left with the task of calculating the trace over the initial state $\rhohat_S$. A close investigation of this expression reveals that the operator acting on $\rhohat_S$ is a displacement operator, $\hatD(\alpha)=e^{\alpha \hatd^\dagger-\alpha^\star \hatd}$, for each Bogoliubov mode with corresponding displacement $\mathrm{i}\gamma_k$.
Due to this and the commutation relations of Bogoliubov operators, $[\hatd_k^\dagger,\hatd_{k'}]=\delta_{k,k'}$, we can write the trace in the last line of Eq.~\eqref{eq:zetamagnus} as the expectation value of a product of displacement operators
\begin{align*}
 \mathrm{trace}
 &= \mathrm{Tr}\left[\prod_k\hatD_k(\mathrm{i}\gamma_k)\rhohat_\beta\right]
 \\
 &=
 \sum_{\{n_k\}}\prod_k \frac{\braket{\hatn_k}^{n_k}}
 	{(1+\braket{\hatn_k})^{n_k+1}}
 	\braket{ \{n_k\} |
 		\prod_{k'}\hatD_{k'}(\mathrm{i}\gamma_{k'}) | \{n_k\}}.
\end{align*}
Here, we have considered that initially the system is in a thermal equilibrium state at inverse temperature $\beta$, so that
$\rhohat_S = \exp(-\beta \hamil_S)/Z$, with the partition function $Z=\mathrm{Tr}[\exp(-\beta \hamil_S)]$, and then used the diagonal representation of the thermal state in the Fock basis.

The action of a displacement operator on a Fock state $\ket{n}$ is to generate a displaced Fock state $\ket{n,\gamma}$. The remaining overlap of two of these states can be expressed by~\citep{displacedfock}
\begin{align*}
&\braket{n,\gamma|m,\alpha}
\\
&=
\braket{\gamma|\alpha}\sqrt{\frac{n!}{m!}}(\gamma^\star-\alpha^\star)^{m-n}L_{n}^{m-n}[(\gamma-\alpha)(\gamma^\star-\alpha^\star)] \:,
\end{align*}
where $L_n^{a}(x)$ are the generalized Laguerre polynomials and  $\braket{\gamma|\alpha}=\exp[-\frac{1}{2}\left(|\gamma|^2+|\alpha|^2-2\gamma^\star\alpha\right)]$
is the overlap of two coherent states.
This enables us to calculate the trace as
\begin{align*}
\mathrm{trace}
=& \sum_{\{n_k\}}\prod_k\frac{\braket{\hat{n}_k}^{n_k}}{(1+\braket{\hat{n}_k})^{n_k+1}} \braket{\{n_k\}|\{n_k\},\{i\gamma_k\}}\\
=&\prod_k \sum_{\{n_k\}}\frac{\braket{\hat{n}_k}^{n_k}}{(1+\braket{\hat{n}_k})^{n_k+1}}e^{-\frac{1}{2}|\gamma_k|^2}L_{n_k}^0(|i\gamma_k|^2) \:.
\end{align*}
%\jmpcom{Variable 'trace' defined two equations above}
This expression can be simplified with the generating function of Laguerre polynomials, $\sum_{n=0}^\infty  t^n L_n(x)=  \frac{1}{1-t} e^{-\frac{tx}{1-t}}$~\citep{ArfkenBook}, which leads to
\begin{align*}
\mathrm{trace}
=
\exp{\left\{ \sum_k\left[ -\frac{1}{2}|\gamma_k|^2
	\coth{\left(\frac{\beta\hbar\omega_k}{2}\right)} \right] \right\}}
\:,
\end{align*}
where we have used $\braket{\hat{n}_k} = 1/(\exp(\beta\hbar\omega_k)-1)$ for a thermal state.
Substituting this result into Eq.~\eqref{eq:zetamagnus} provides Eq.~\eqref{eq:decfuncbogo}.

%%%%%%%%%%%%%%%%%%%%%%%%%%%%%%%%%%%%%%%%%%%%%%%%%%%%%%%%%%%%%%%%%%%%
%%%%%%%%%%%%%%%%%%%%%%%%%%%%%%%%%%%%%%%%%%%%%%%%%%%%%%%%%%%%%%%%%%%%

\section{Calculation of the variance}
\label{app:errors}
We show how to estimate the uncertainty in the measurement of $\mathrm{Re}(\zeta(t))$ due to the projection noise on the measurement of the state of the qubits.
In this way, we determine the noise which has to be added to the calculated values of the coherence function to simulate the outcome of experiments.

In accordance with Eq.~\eqref{eq:Repart},
\begin{align}
 \mathrm{Re}(\zeta(t))
 &=\frac{1}{2}
 \braket{ \hat\sigma_x\otimes\hat\sigma_x 
  -\hat\sigma_y\otimes\hat\sigma_y} \:,
 \label{eq:RepartApp}
\end{align}
the real part of the coherence function can be determined by measuring the expectation value of a combination of Pauli matrices on the state of the qubits. Hence, we start by calculating the variance associated with this expectation value. Introducing the shorthand notation $\hat\sigma_{xx}=
\hat\sigma_x\otimes\hat\sigma_x$, and similarly for $\hat\sigma_{yy}$ and $\hat\sigma_{zz}$, we have
\begin{align*}
 \mathrm{Var}(\hat\sigma_{xx}-\hat\sigma_{yy})
 = \braket{(\hat\sigma_{xx}-\hat\sigma_{yy})^2}
  -\braket{\hat\sigma_{xx}-\hat\sigma_{yy}}^2
 \:.
\end{align*}
The last term is directly related to the coherence function $\braket{\hat\sigma_{xx}-\hat\sigma_{yy}}^2=4\mathrm{Re}(\zeta(t))^2$, whereas the first can be calculated as
\begin{align}
 \langle(\hat\sigma_{xx} -\hat\sigma_{yy})^2\rangle
 &=
 \braket{\hat\sigma_{xx}^2+\hat\sigma_{yy}^2
 		-\hat\sigma_{xx}\hat\sigma_{yy} -\hat\sigma_{yy}\hat\sigma_{xx}}
 \nonumber 
 \\
 &=
 2\braket{\identmat_4 +\hat\sigma_{zz}}
 \:
 \label{eq:variance}
\end{align}
where $\identmat_4$ is the $4\times4$ identity matrix. In the last line, we have used that the Pauli matrices fulfill the algebraic relation $\hat\sigma_a \, \hat\sigma_b  = \delta_{ab}\identmat_2 + \mathrm{i}\, \sum_{c=x,y,z} \epsilon_{abc}\, \hat\sigma_c$.
We observe that the right-hand side of Eq.~\eqref{eq:variance} is a diagonal matrix. Since the time evolution does not affect the diagonal elements, we can evaluate this expectation value over the initial Bell state, resulting in $\langle(\hat\sigma_{xx} -\hat\sigma_{yy})^2\rangle = 4$. Thus,
\begin{align*}
 \mathrm{Var}(\hat\sigma_{xx}-\hat\sigma_{yy})
 &= 4 \left[ 1-\mathrm{Re}(\zeta(t))^2 \right] \:.
\end{align*}
Substituting this into Eq.~\eqref{eq:RepartApp}, it follows that the variance of the real part of the coherence function is connected to the function itself via
\begin{align*}
 \mathrm{Var}(\mathrm{Re}(\zeta(t))
 &=
 \frac{1}{4}\mathrm{Var}(\hat\sigma_{xx}-\hat\sigma_{yy} )
 = 1-\mathrm{Re}(\zeta(t))^2 \:.
\end{align*}
For the error on the imaginary part of the coherence function, the calculation is analogous.

Having determined the variances of the real and imaginary parts of $\zeta(t)$, we simulate the uncertainty in experiments by adding Gaussian noise of zero mean and standard deviations  $\sigma_\mathrm{Re}=\sqrt{1-\mathrm{Re}(\zeta(t))^2}$ and $\sigma_\mathrm{Im}=\sqrt{1-\mathrm{Im}(\zeta(t))^2}$ to the real and imaginary parts, respectively.

%\bibliography{library}
\bibliography{twoprobes-biblio} % bibexport -o twoprobes-biblio.bib twoprobes-rev-v1.aux

\end{document}